\documentstyle[11pt,paspconf,epsf]{article}
\def\edcomment#1{\iffalse\marginpar{\raggedright\sl#1\/}\else\relax\fi}
\reversemarginpar

\begin{document}
\title{New HI Features of the Magellanic System}
\author{M. E. Putman}
\affil{Mount Stromlo and Siding Spring Observatories, ANU, ACT 2611, Australia}
\author{B. K. Gibson}
\affil{CASA, University of Colorado, Boulder, CO 80309}
\author{L. Staveley-Smith}
\affil{Australia Telescope National Facility, CSIRO, NSW 2121, Australia}

\begin{abstract}

 The first results from the HI Parkes All-Sky Survey (HIPASS) 
provide a spectacular view of the global HI
distribution in the vicinity of the Magellanic Clouds and the 
southern Milky Way.   A 2400 square degree mosaic around the South 
Celestial Pole (SCP) reveals the existence of a narrow, 
continuous counter-stream which 'leads' the direction of motion 
of the Clouds, i.e. opposite in direction to the Stream.  
This strongly supports the gravitational model for the Stream 
in which the leading and trailing streams are tidally torn 
from the body of the Magellanic Clouds. 
The data also reveal additional tidal features in the Bridge region which
appear to emanate from the LMC, and a distinct spiral structure within
the LMC itself.
  
\end{abstract}


\section{Introduction \& Observations}

The gas distribution in the Magellanic Clouds, unlike the stellar component,
 immediately suggests that
the Clouds are not independent entities, but part of a dynamically interacting 
system.  Recent observations taken as part of the HI Parkes All-Sky Survey (HIPASS)
reveal several exciting new features of this system.  The single dish 21-cm
observations allow us to examine the large scale structure of the LMC,
the overall spatial and velocity distribution of the SMC and Bridge gas and 
resolve the long-standing controversy about the mechanism responsible for
the formation of the Magellanic Stream.

HIPASS is a survey for extragalactic neutral hydrogen in the southern sky ($\delta\le +2^\circ$),
which has also proven useful for studies
of the local gas distribution (i.e. -400 - +400 km s$^{-1}$).
The survey is being completed by actively scanning the 64-m Parkes telescope,
equipped with the 13 feed-horn multibeam receiver, across the sky in declination 
(Staveley-Smith 1997).
The entire southern sky has already been scanned once with a spacing on
the sky of $\sim7^{\prime}$ and a sensitivity of 20 mK (1$\sigma$).  The 
complete survey will consist
of five separate scans, resulting in sub-Nyquist sampling and a resulting sensitivity 
of 9 mK (1$\sigma$).  The channel spacing is 13.2 km s$^{-1}$, with hanning smoothing resulting
in a final resolution of 26.4 km s$^{-1}$.
The survey's unbiased coverage
of the southern sky and dense spatial sampling provide us with the 
ideal tools to reveal
previously unidentified HI structure about the Magellanic Clouds.

\section{Results}
The data presented here are from the first scan of HIPASS data.
Starting with the LMC (Figure 1), we immediately see the advantages
of fully sampled single dish observations.  Figure 1 highlights
the distinct spiral structure of the LMC and also depicts the  
diffuse HI edges on all sides of the Cloud.  This contradicts the suggestion 
that the gas is being compressed as the 
LMC moves through the halo of the Milky Way (e.g. de Boer et al 1998); 
however, there is still a steep gradient in the east and the diffuse nature 
of the outer gas may be due to differential rotation.

The LMC appears to have 3 spiral ``arms'' (Figure 1) and one 
southern extension which looks like another thin spiral arm until
its relationship to the Magellanic Bridge gas is examined.
Figure 2 shows the LMC+Bridge+SMC as seen by HIPASS.  In this picture
the LMC's southern extension is clearly seen as a tidal feature which
merges with gas from the SMC, forming the Bridge.  This figure also shows the beginning of
the Magellanic Stream which continues for another $100^\circ$ to
the north (Mathewson et al 1974).  The clumpy nature of the Stream is immediately apparent, even in this
limited region.

In addition to the Magellanic Stream, figure 2 also
shows the beginning of a thin southern feature emanating from the SMC+Bridge 
region $(\ell,b = 292^\circ, -32^\circ)$.
Moving out to a 2400 square degree mosaic about
the South Celestial Pole (Figure 3), we see that this feature continues 
``towards'' the Galactic Plane, on the leading side of the Clouds.
This extension is at
very high positive velocities and is clearly separate from Galactic emission
(see Putman et al 1998).
The leading feature continues from the Clouds both spatially and in velocity,
 to $b \approx -8^\circ$ and V$_{LSR}$ = 350 km s$^{-1}$.

This leading extension is exactly the type of feature predicted to form during a
tidal interaction between the Clouds and the Milky Way (e.g. Gardiner \&
Noguchi 1996).
There are two principal interaction scenarios which attempt to explain the 
presence of the Magellanic Stream.
One uses gravitational tides to pull out leading and trailing 
streams from the Clouds; the other employs ram-pressure
stripping to rip the Stream out of the Clouds as they pass through
an extended Galactic halo.  The argument against tidal models (and for the
ram-pressure models) is
largely based on two observational features:  no leading stream has
been detected, and there are no stars observed in the Stream (stars should also 
be affected by tidal forces).
The leading feature of figure 3 removes the first discriminant against tidal
models.  The second issue, not finding stars in the Stream, is actually
quite complex.  It is common to find systems with strong HI tidal features and
no optical counterpart (e.g. the M81 system (Yun et al 1994)) or HI 
streams offset from the optical streams (e.g. Smith et al 1997).
It may be that we have not looked in the right place for this optical counterpart
to the Magellanic Stream, or, considering that the Stream is believed 
to be $\sim 1.5$ Gyr old, we may not be looking for the right age of
stars (Majewski et al 1998).  In addition, it is probable that
the gas in the Clouds was originally more extended than the stars.
Thus, the discovery of the Leading Arm is the strongest support in 25 years for
a past tidal interaction between the Magellanic Clouds and the Milky Way.
 
There are several ways in which the tidal interaction scenario can be confirmed.
HIPASS will continue to map the HI about the Magellanic 
System, allowing us to search for a continuation of the Leading Arm and other
Magellanic-related features.  A preliminary look at the data on the other
side of the Galactic Plane suggests that the arm veers to the east and 
continues to $b \approx 30^\circ$.
We will also increase the sensitivity of our present data by adding more 
multibeam scans and increase our velocity resolution by observing these features
with the narrow-band facility.  
Emission line observations and metallicity determinations will also
prove useful for constraining the dominant ionization mechanism in the Leading
Arm region and confirming the Magellanic origin.


\acknowledgments
We thank the multibeam working group for their help in the data collection
and development of the instrumentation and software.

\begin{figure}[h]
\vspace{0.2in}
\caption{HIPASS map of the LMC.  Note the distinct spiral
arm in the south and the two stubs in the north.}  
\end{figure}

\begin{figure}[h]
\vspace{0.2in}
\caption{Brightness temperature map of the LMC+SMC+Bridge region including 
$V_{LSR}$ = 85 - 400 km s$^{-1}$.  Contours are a geometric progression from 1.3 - 83 K.}
\end{figure}

\begin{figure}[h]
\vspace{0.2in}
\caption{Brightness temperature map of the South Celestial Pole including
$V_{LSR} = 85 - 400$ km s$^{-1}$. }
\end{figure}

\end{document}